\documentstyle[preprint,aps,prc]{revtex}
\input{psfig}
\tightenlines

\begin{document}

\title{Pion and photon couplings of $N^*$ resonances \\
       from scattering on the proton}

\author{A.Yu. Korchin}
\address{Department of Subatomic and Radiation Physics,
         University of Gent, \\ Proeftuinstraat 86, B-9000 Gent, Belgium,}
\address{and}
\address{National Science Center
         `Kharkov Institute of Physics and Technology,'
         \\ 310108 Kharkov, Ukraine}

\author{O. Scholten and R.G.E. Timmermans}
\address{Kernfysisch Versneller Instituut, University of Groningen, \\
         Zernikelaan 25, 9747 AA Groningen, The Netherlands}


\maketitle

\begin{abstract}
Results of a relativistic model for pion- and photon-induced
reactions on the proton are presented. The model is crossing
symmetric and gauge invariant. The nucleon resonances $P_{11}(1440)$,
$P_{11}(1710)$,
$D_{13}(1520)$, $S_{11}(1535)$, $S_{11}(1650)$, $P_{33}(1232)$,
$P_{33}(1600)$, $S_{31}(1620)$, and $D_{33}(1700)$ have been
included explicitly in the calculation. Unitarity within the
channel space $\pi N\oplus\gamma N$ below the two-pion production
threshold has been achieved by using the $K$-matrix approach.

Strong and electromagnetic coupling parameters of the $N^*$
resonances have been determined from a fit to the pion-nucleon
phase shifts, pion-photoproduction multipoles and
Compton-scattering cross sections. The model is shown to
describe simultaneously most of the available data. Results for
the electric and magnetic polarizabilities of the nucleon
are presented.
\end{abstract}

In recent years effective Lagrangian models have been applied
to pion-nucleon scattering
$\pi N\rightarrow \pi N$~\cite{Pea91,Lee91,Gou94,Gro93,Pas97},
pion photoproduction
$\gamma N\rightarrow \pi N$~\cite{Dav91,Noz90,Gar93,Sur96,Feu97,Van95},
and Compton scattering $\gamma N\rightarrow\gamma N$~\cite{Pas95}
on the nucleon. In these
calculations where nucleon resonances are explicitly included
the interference between the resonant and background contributions
plays an important role. From pion photoproduction
calculations~\cite{Pec69,Ols75,Dav91,Van95}
it is known that the unitarity constraint is important to obtain the
correct interference. In most calculations unitarity was imposed by
using Watson's theorem in the $\Delta(1232)$ resonance channel. In
Refs.~\cite{Pea91,Lee91,Gro93,Pas97} $\pi N$ rescattering effects
were studied in the framework of three-dimensional equations,
which also guarantee (two-body) unitarity.

In the case of Compton scattering strong constraints on the amplitude
follow from unitarity and causality. These constraints are usually
formulated in terms of fixed-$t$ dispersion
relations~\cite{Bal60,Ach77,Gui79,Lvo79} by relating
the imaginary part of the amplitude via the optical theorem to
the pion-photoproduction cross section. Although such a dispersion
relation approach has many advantages, unfortunately the link with
the decay properties of the nucleon resonances is rather obscure.
In particular, it is difficult to extract parameters of the
electromagnetic decays of resonances $N^*\rightarrow N\gamma$.

It is the purpose of this Letter to study all three above processes
in an unified approach based on an effective Lagrangian model and
unitarity, and to extract the pion and photon couplings of the
$N^*$ resonances below $\sqrt{s}\simeq 1.7$ GeV.
The advantage is that the different processes are calculated
consistently, therefore putting stronger constraints on the
parameters. In an earlier paper~\cite{Sch96} a similar approach was
used in the $\Delta(1232)$ resonance region. Here, energies up to
$\sqrt{s}\simeq 1.7$ GeV are studied and correspondingly the heavier
nucleon resonances are also explicitly included. The model is fully
relativistic, crossing symmetric, and for pion photoproduction and
Compton scattering also gauge invariant.

A convenient approach for imposing the unitarity constraint
is to work with the $K$-matrix formalism. It results from the
Bethe-Salpeter (BS) equation in the approximation where only the
discontinuity part of the loop integrals is kept. In other words,
the particles forming loops are taken to be on the mass shell. The
$S$ matrix in this approach is unitary and symmetric provided the
$K$ matrix is taken to be real and hermitian. Thus choosing the
tree-level diagrams as the $K$ matrix ensures (two-body) unitarity
within the model space $\pi N\oplus\gamma N$. Two- and multi-pion
production channels become important at energies $\sqrt{s}\simeq 1.2$
GeV. These additional decay channels can be taken into account
approximately by including the corresponding width in the resonance
propagator.

The reaction $T$ matrix in the channel space labeled by the
index $c=\pi N$ or $\gamma N$ is written schematically as
\begin{equation}
   T_{c'c} = K_{c'c} + {i\over (2\pi)^4} \sum _{c''}K_{c'c''}
             \tilde{G}_{c''c''} T_{c''c} \ , \label{eq:1}
\end{equation}
with the appropriate two-body propagator $\tilde{G}$.
The minimal requirement for this $\tilde{G}$ is the correct
discontinuity across the cut in the complex energy plane $W=\sqrt{s}$.
This guarantees the unitarity of the $S$ matrix when the kernel $K$
is hermitian. The discontinuity can be obtained by applying the
Cutkosky rules. Since the $S$ matrix in our approach is unitary,
this implies that Watson's theorem for pion photoproduction as
well as the optical theorem for Compton scattering~\cite{Pfe74}
are satisfied.

The $K$ matrix is calculated from the tree-level amplitude for
pion-nucleon scattering, pion photoproduction and Compton scattering.
This tree-level amplitude contains the following contributions: In
the $s$- and $u$-channel exchange of the nucleon and the (four-star)
$N^*$ resonances~\cite{PDG}: $P_{11}(1440)$, $D_{13}(1520)$,
$S_{11}(1535)$, $S_{11}(1650)$, $P_{33}(1232)$, $S_{31}(1650)$,
and $D_{33}(1700)$. We also include the (three-star) $P_{11}(1710)$
listed by the PDG~\cite{PDG}, although the PDG and the SM95 solution
of the Virginia Tech $\pi N$ PWA~\cite{Arn95} differ significantly
for the pole position, as well as the three-star $P_{33}(1600)$
resonance, for which also the pole position is uncertain. We do
not include the $D_{13}(1700)$ listed by the PDG as a three-star
resonance, but not seen in SM95. Also not taken into account are
spin-5/2 resonances, in particular the $D_{15}(1675)$ and
$F_{15}(1680)$.

Additionally included in the tree-level amplitude are the
$t$-channel exchange of $\sigma(760)$ and $\varrho(770)$ mesons for
$\pi N$ scattering, $\pi(140)$, $\varrho(770)$, and $\omega(781)$
mesons for pion photoproduction, and $\pi^0(135)$ and $\sigma(760)$
mesons for Compton scattering. The $\sigma(760)$ could be seen
as a rough manner to take into account a number of isoscalar
$t$-channel exchanges (in particular, isoscalar two-pion,
$\varepsilon(760)$, and ``pomeron'' exchange).

For the spin-3/2 resonances we take the most general coupling
to the $\pi N$ and $\gamma N$ channels, including off-shell
contributions and the full Rarita-Schwinger propagator. In this
paper, we consider energies up to $\sqrt{s}\simeq 1.7$ GeV
and assume that the description in terms of nucleon resonances
and $t$-channel meson exchanges is adequate.
For higher energies, when one enters the Regge regime, one
faces acute problems such as issues related to reggeizing the
exchanges, duality, and possible double counting. (A recent
model~\cite{Gui97} is an example of reggeizing $t$-channel exchanges
for pion and kaon photoproduction at energies $E_\gamma>4$ GeV.)

Above the two-pion production threshold at $\sqrt{s}=1.22$ GeV
the nucleon resonances acquire an additional two-pion decay
width. For the $\Delta(1232)$ resonance this additional width
is negligible, while for the heavier resonances we assume its
energy dependence to be of the form
\begin{equation}
   \Gamma(W) = \Gamma_0\,{4x\over (1+x)^2} \,
                  \theta(W^2-(m+2m_\pi)^2) \ , \label{eq:2}
\end{equation}
where
\begin{equation}
           x = \left( {W^2-(m+2m_\pi)^2\over
               M_r^2-(m+2m_\pi)^2 } \right)^3 \ . \label{eq:3}
\end{equation}
Here, $M_r$ is the resonance mass. The energy dependence is chosen
such as to approximately take a three-particle phase space into
account. For the $S_{11}(1535)$ resonance the additional width
comes mainly from the $N\eta$ decay channel; in this case, we
take $m_\eta$ instead of $2m_\pi$ in Eqs.~(\ref{eq:2}) and
(\ref{eq:3}).
The parameter $\Gamma_0$ is chosen for each resonance
equal to the decay width outside the $\pi N$ channel. This
additional width is included in the resonance propagators by
changing $M_r$ into $M_r-i\,\Gamma(W)/2$. The inclusion of this
additional width gives an imaginary part to the $K$ matrix, and
accounts for all channels not included in the space
$\pi N\oplus\gamma N$.

Concluding the description of the model, we mention that the above
amplitude (for pion photoproduction and Compton scattering) obeys
gauge invariance provided the $K$ matrix is constructed to be gauge
invariant. Where there are no form factors in the Born diagrams
gauge invariance is satisfied, while the resonance contributions are
gauge invariant by themselves also when form factors are included.
However, when form factors are included in the $\pi N\!N$ and/or
$\gamma N\!N$ vertices in the Born diagrams additional terms are
needed in the pion-photoproduction and Compton amplitudes to fulfill
gauge invariance. To construct these terms we have used minimal
substitution, following the method of Ref.~\cite{Oht89}.
In particular, the additional terms for pion
photoproduction reproduce the Kroll-Ruderman term when the form
factors are put equal to unity. We have checked explicitly
that the resulting amplitudes are indeed gauge invariant
numerically.

\begin{figure}[1]
\vspace*{6cm}
\psfig{figure=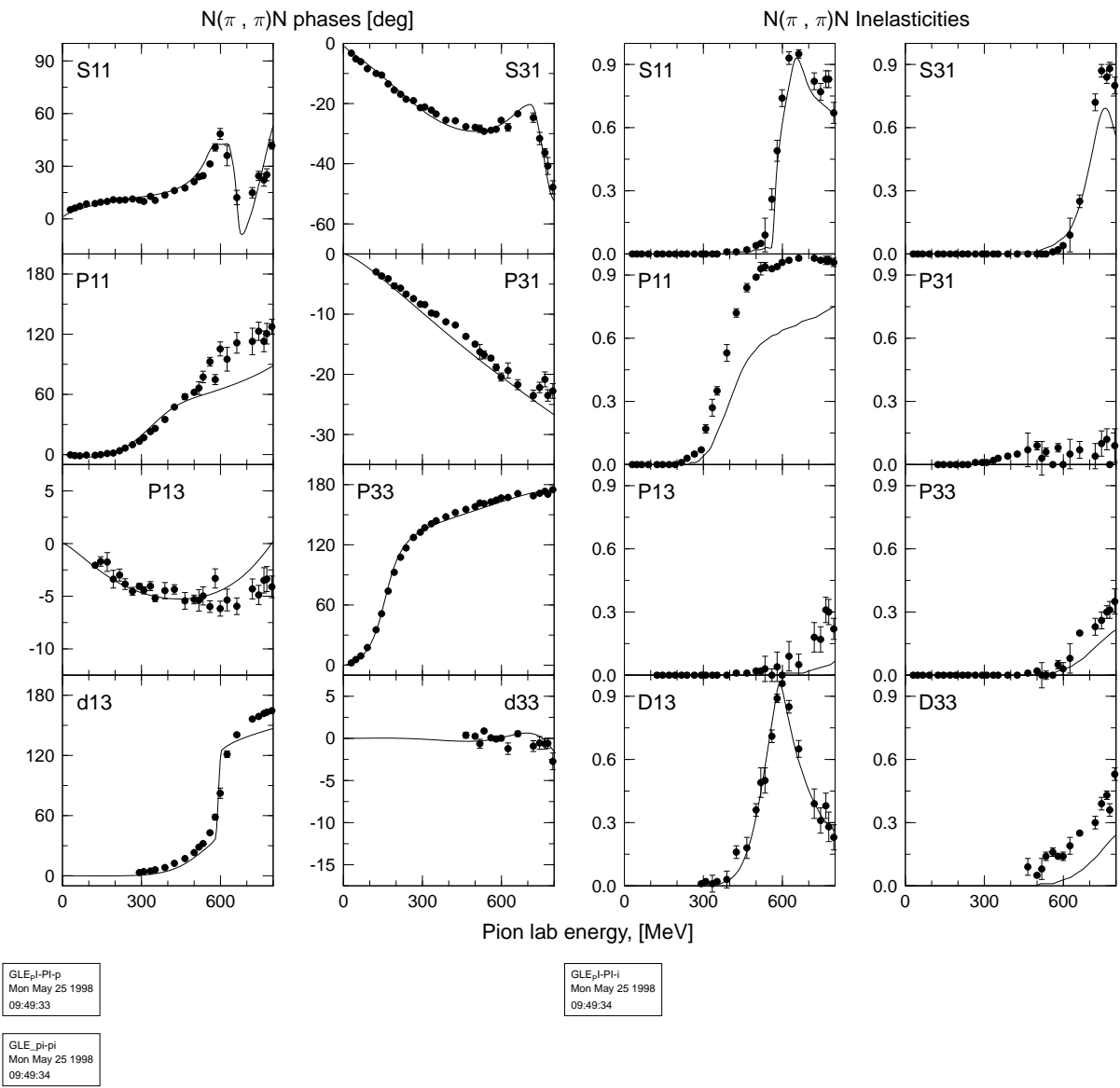,height=5.1in}
\caption{Pion-nucleon $S$-, $P$-, and $D$-wave phase shifts
         and inelasticities for $I=1/2$ and $I=3/2$.
         The curves are our model results, the points
         are taken from the Virginia Tech single-energy
         partial-wave analysis~\protect\cite{SAID}.}
\label{fig:1}
\end{figure}

\begin{figure}[2]
\vspace*{7.2cm}
\psfig{figure=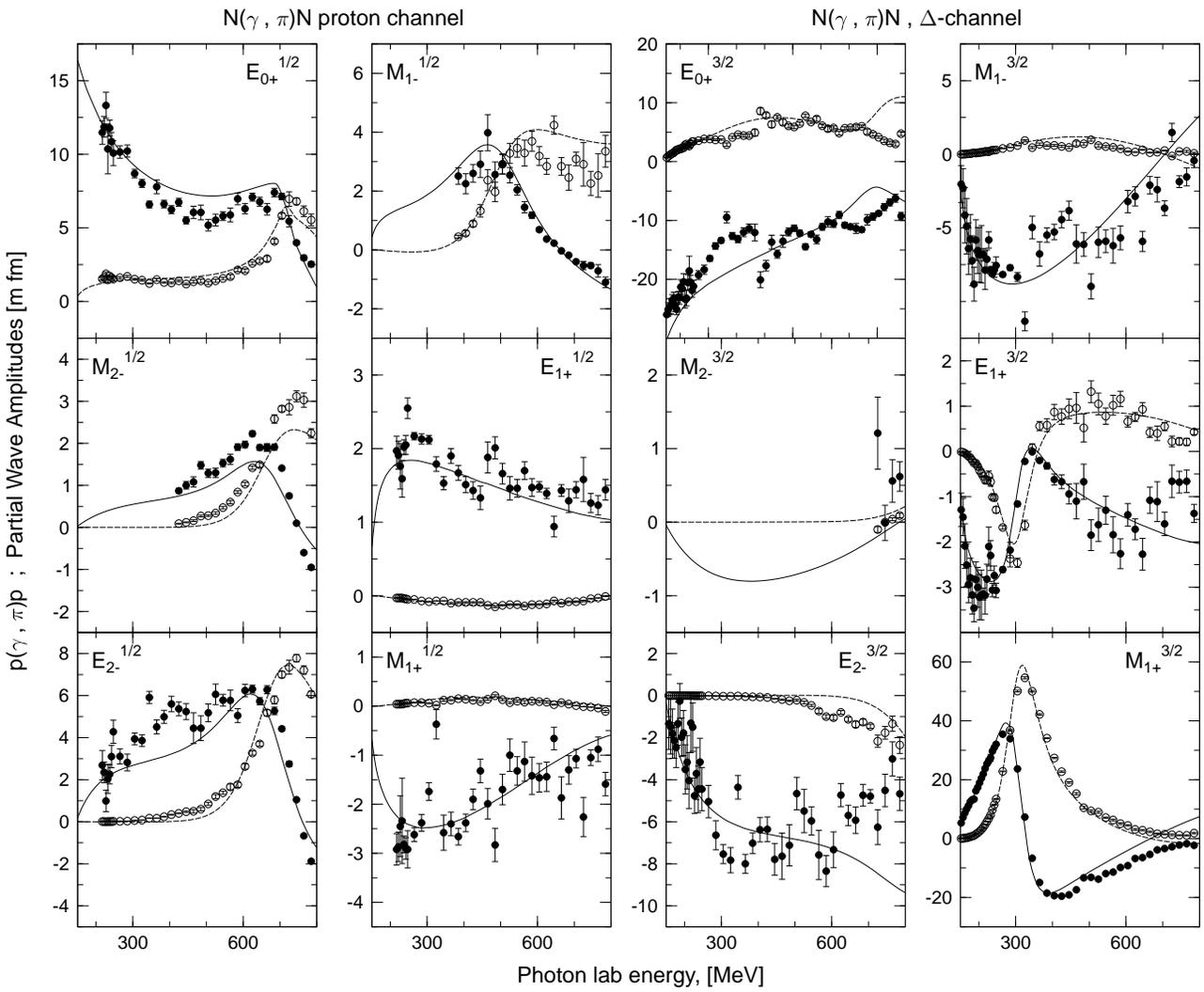,height=5.3in}
\caption{Electromagnetic multipoles in mfm. The solid (dashed)
         curves are the real (imaginary) parts of the multipoles.
         The points are taken from the Virginia Tech single-energy
         partial-wave analysis~\protect\cite{SAID}.}
\label{fig:2}
\end{figure}

\begin{figure}[3]
\vspace*{6.5cm}
\psfig{figure=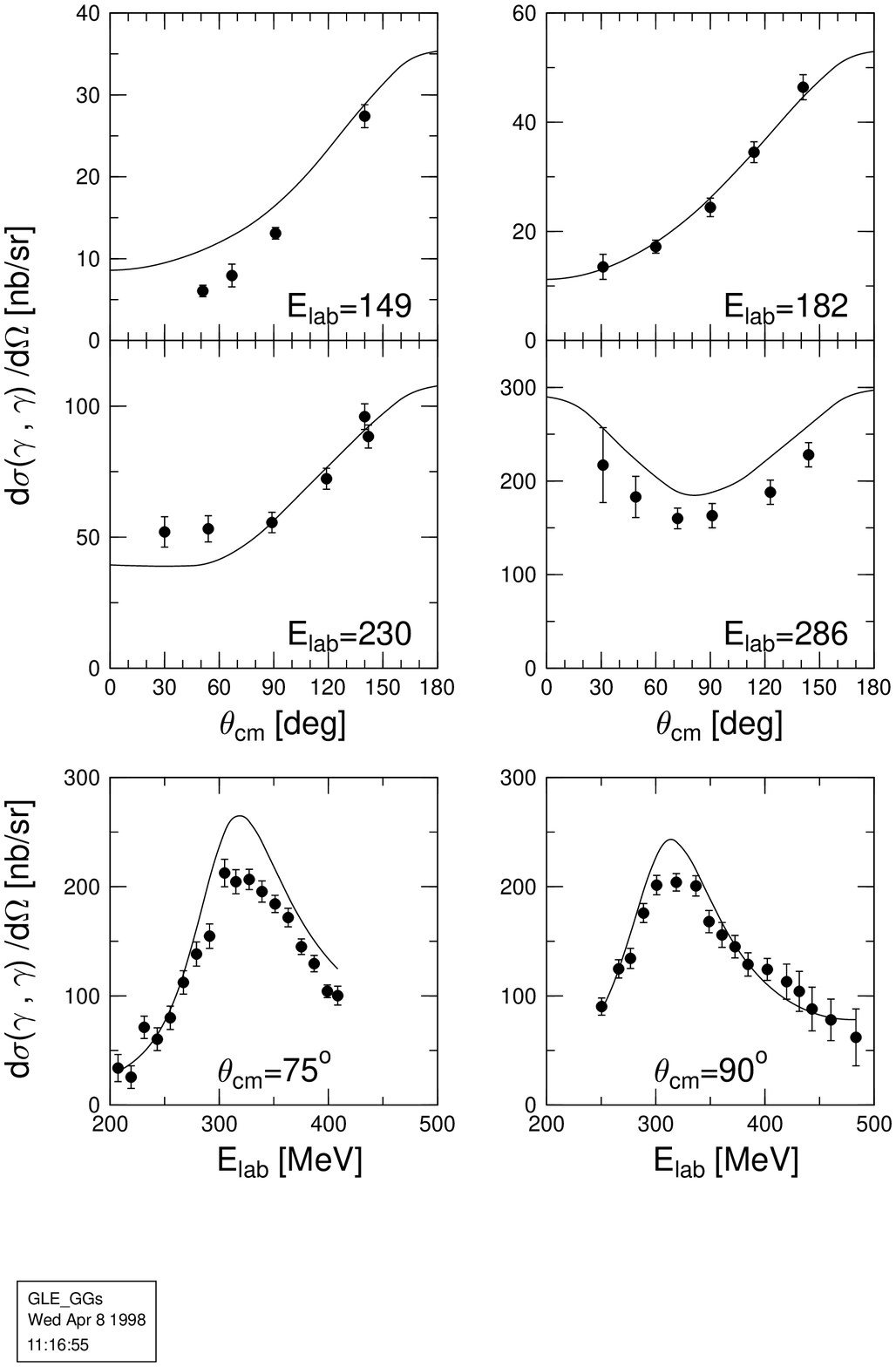,height=5.3in}
\caption{Differential cross sections for proton Compton scattering
         (top and middle) and cross sections at $\theta_{cm}=75^\circ$
         and $\theta_{cm}=90^\circ$ (bottom). The data points
         are from~\protect\cite{Hal93} (angular distributions)
         and from~\protect\cite{Mol96,Pei96,Hun97} (at fixed photon
         angles). The curves are the results of the present model.}
\label{fig:3}
\end{figure}

The parameters in the model have been fitted to the Virginia Tech
single-energy partial-wave amplitudes~\cite{SAID} of Arndt and
collaborators. The pion-nucleon $S$-, $P$-, and $D$-wave phase
shifts and inelasticities for isospin $I=1/2$ and $I=3/2$ are
fitted up to $T_\pi=820$ MeV pion lab energy. The corresponding
pion-photoproduction multipoles are fitted up to $E_\gamma=805$ MeV
photon lab energy. Finally, Compton scattering cross sections are
also included in the fit, in particular, recent cross sections from
Saskatoon and Mainz ~\cite{Hal93,Bla96,Mol96,Pei96,Hun97}.
The limited number of free parameters (see below and
Tables~\ref{tab:1} and~\ref{tab:2}) are strongly constrained by
such a combined fit. Anecdotal results of the fit are shown in
Figs.~\ref{fig:1}--\ref{fig:3}.
In the following, we will focus attention to some particular
details of our results, discussing in turn pion-nucleon scattering,
pion photoproduction, and Compton scattering.

The pion-nucleon phase shifts are well described, although deviations
start to show up at the upper limit of the energy range considered.
The inclusion of an additional width in the propagator accounts
rather well for the observed inelasticities for pion-nucleon
scattering. However, the peculiar ``saturation'' of the inelasticity
in some partial waves, in particular $P_{11}$, could not be reproduced
completely.
The $P_{11}$ partial wave has more peculiar features. An attempt to
describe it in our approach with one ``Roper'' resonance situated
at approximately 1.44 GeV failed. We find that the background
contribution to this $P_{11}$ phase shift due to $t$- and $u$-channel
processes is rather large, about 60$^\circ$ at 1.5 GeV. When a
resonance at 1.44 GeV is added to such a background, one obtains a
phase shift that increases rapidly to a value larger than 180$^\circ$.
An acceptable description of the $P_{11}$ phase shift could be obtained
by placing an additional resonance at a higher energy of $\sqrt{s}=1.71$
GeV.

As mentioned, off-shell form factors for the nucleon resonances have
been introduced in the model. We have not investigated the importance
of the detailed dependence on the invariant mass and merely adjusted
the cutoff. The form factor is normalized to unity at the resonance and
suppresses the resonance contribution at higher and lower invariant
masses.
This
choice thus leads to a suppression of in particular the $u$-channel
exchange diagrams. This could be justified in the cloudy-bag
model~\cite{The80} where an $N^*$ resonance is considered
as a mixture of the three-quark and $\pi N$ scattering states
(see also the discussion in Ref.~\cite{Gar93}).
The cutoff form factors are taken to depend on the invariant mass
of the intermediate $N$ or $N^*$, that is, on the Mandelstam
variable $s$ for $s$-channel exchanges and on $u$ for $u$-channel
exchanges, {\it viz.}
\[ F(x) = \frac{\Lambda^4}{\Lambda^4+(x-M_r^2)^2} \ , \]
where $M_r$ is the nucleon or resonance mass, and $x$ equals
$s$ or $u$.
In this way the crossing-symmetry properties of the
amplitudes remain the same as for the case without form factors.
The cutoff mass of the form factors has been set to 1.2 GeV.
The results are not very sensitive to the precise value.

The $\pi N^*N$ coupling for spin-3/2 resonances contains two
parameters, the strength $g$ and the off-shell parameter $z_\pi$,
which determines the coupling to the spin-1/2 sector of the
Rarita-Schwinger propagator~\cite{Pec69,Ben89}. In the
following, we will also use the parameter $a_\pi =-(0.5+z_\pi)$.
It is known that this off-shell parameter is important for a good
description of pion photoproduction~\cite{Dav91,Feu97} and Compton
scattering~\cite{Pas95}. We found that the main effect of
adjusting $z_\pi$ is an overall shift in the spin-1/2 channel with
the same isospin (and opposite parity) as the resonance channel. The
influence of the off-shell parameter for the $P_{33}(1232)$ resonance
on the $S_{31}$ partial wave and of the off-shell parameter for
the $D_{13}(1520)$ on the $S_{11}$ and $P_{11}$ partial waves
are especially pronounced. The value of the off-shell parameter
for the $\Delta(1232)$ favored by the fit, $a_\pi=-0.265$, is close
to the ``Peccei value''~\cite{Pec69,Ben89} $z_\pi=a_\pi=-0.25$.

In this connection, we mention the tree-level calculations
of pion photoproduction of Refs.~\cite{Gar93,Feu97}. In
particular, in Ref.~\cite{Gar93} $u$-channel contributions had
to be suppressed in order to get a reasonable agreement with
the pion-photoproduction data. However, the off-shell coupling
of all the spin-3/2 resonances were set to $z_\pi=-0.25$ from
the beginning. In Ref.~\cite{Feu97}, on the other hand,
these parameters were determined from the data; the resonance
parameters extracted were quite different from those of
Ref.~\cite{Gar93}. The authors of~\cite{Feu97} also used a cutoff
only for the $u$-channel diagrams, different from our prescription.

For the $\pi N^*N$ couplings, we assume a derivative coupling
of the pion. For the $\pi N\!N$ case, {\it e.g.}, we take
the standard pseudovector Lagrangian.
For the $D_{13}(1520)$ resonance a very specific coupling was
required. The general Lagrangian is, with $R=D_{13}$,
\begin{equation}
  {\cal L}_{\pi N\!R} =
    \frac{f_{\pi N\!R}}{m_\pi}\,\overline{N}
    \left[(1-\chi) \frac{i\,\slash\hspace{-2.5mm}{\partial}}{M+M_R}
    + \chi\right] (g_{\alpha\beta}+a_\pi\gamma_\alpha\gamma_\beta) \,
    \partial^\alpha\vec{\pi}\cdot\vec{\tau}\:
    \gamma_5\, R^\beta + \;{\rm h.c.}
\end{equation}
For $\chi=0$, a strong coupling to the ``negative-energy'' components
of the nucleon results, leading to an undesirable resonance-like peak
in the $P_{13}$ partial wave. For nonzero and increasing $\chi$ this
coupling becomes more-and-more suppressed, and a reasonable
description of the $P_{13}$ phase shift is possible for $\chi=0.5$.

In Table~\ref{tab:1}, we list the results for the different
resonances, their pole positions, one-pion couplings, and
additional (two- and multi-pion) decay widths. For the spin-3/2
resonances we also give the off-shell parameters $z_\pi$ and
$z_{1,\gamma}, z_{2,\gamma}$. The one-pion couplings
are in general in good agreement with those listed in the
Tables of the Particle Data Group, although for the $S_{11}(1650)$
and $S_{31}(1620)$ they are on the high side.

The pion-nucleon coupling constant is fixed~\cite{Tim97} to
$g_{\pi N\!N}=13.02$. For the coupling constants of the $\varrho(770)$
we get the following results: $g_{\pi\pi\varrho}=6.07$ is
taken from the two-pion decay width. In the fit we get
then for the couplings to the
nucleon: $g_{\varrho N\!N}=4.12$ and $\kappa_\rho=1.79$. The fit
is only sensitive to the product $g_{\pi\pi\varrho}g_{\varrho N\!N}$.
When we assume, following Sakurai, universal coupling of the
vector mesons to the isospin current, we have
$g_{\pi\pi\varrho}=g_{\varrho N\!N}=5.00$. Our result for
$\kappa_\varrho$ is significantly lower than the prediction
$\kappa_\varrho=3.71$ from vector dominance of the nucleon
electromagnetic form factors. The $\omega(781)$ coupling
constant $g_{N\!N\omega}$ follows from assuming $SU(3)$ symmetry,
ideal mixing of the vector mesons, and Sakurai's universality
as $g_{N\!N\omega}=3/2\,g_{N\!N\varrho}=7.50$. Vector dominance
gives $\kappa_\omega=-0.12$. The $\sigma(760)$ couplings
to the nucleon and pion are defined as in Ref.~\cite{Gou94},
{\it viz.}
\begin{equation}
   {\cal L}_{N\!N\sigma}   = -g_{N\!N\sigma}\,\overline{N}\!N\sigma
   \ ,  \;\;\;\;
   {\cal L}_{\pi\pi\sigma} = -g_{\pi\pi\sigma}\,m_\pi\,
                             \vec{\pi}\cdot\vec{\pi}\,\sigma \ ,
\end{equation}
where only the product $g_{\pi\pi\sigma}g_{N\!N\sigma}$
appears in the $\pi N$ scattering amplitude.
We get $g_{\pi\pi\sigma}g_{N\!N\sigma}=4.7$.

In general, a good fit of the electromagnetic multipoles in pion
photoproduction has been obtained. The resulting photon couplings
to the different resonances can be found in Table~\ref{tab:2}.
The meson-photon coupling constants are determined from the decay
widths~\cite{Gar93}:
    $g_{\varrho^\pm\pi\gamma}=0.103$,
    $g_{\varrho^0\pi\gamma}=0.131$, and
    $g_{\omega\pi\gamma}=0.313$.

It is found from fitting the Compton scattering cross sections
that the parameters of the $\Delta N\gamma$ vertex are
determined rather accurately by these data. In particular, the
near-vanishing of the cross section at forward angles close to the
pion-production threshold is due to a cancellation between nucleon
and $\Delta(1232)$ contributions. At large angles the $\sigma(760)$
meson gives an important contribution to the cross section, while
the $t$-channel contributions vanish at forward angles where $t=0$.
The $\pi^0$-exchange contribution is fixed without free
parameters from the axial-anomaly Lagrangian~\cite{Wei96}
\begin{equation}
   {\cal L}_{\pi^0\gamma\gamma} =
        \frac{e^2g_{\pi^0\gamma\gamma}}{2m_{\pi^0}}
        \,\varepsilon_{\mu\nu\alpha\beta}
        \,\partial^\mu A^\nu\,\partial^\alpha A^\beta \pi^0 \ ,
\end{equation}
where $g_{\pi^0\gamma\gamma}=N_c m_{\pi^0}/12\pi^2f_\pi=0.038$
with $N_c =3$ and $f_\pi=92.4$ MeV.
The photon coupling to the $\sigma$-meson is
\begin{equation}
   {\cal L}_{\sigma\gamma\gamma} = \frac{e^2g_{\sigma\gamma\gamma}}
                   {2 m_\sigma}\,\partial^\mu A^\nu
       (\partial_\mu A_\nu-\partial_\nu A_\mu)\,\sigma \ ,
\end{equation}
where $g_{\sigma\gamma\gamma}$ is determined
accurately from the Compton data; we found
$g_{\sigma\gamma\gamma}g_{N\!N\sigma}=-5.65$, which for 
$g_{\sigma\gamma\gamma}= -0.565$ corresponds to the
$\sigma\rightarrow\gamma\gamma$ decay width $\Gamma=10.15$ keV.

Apart from the $P_{33}(1232)$ and $D_{13}(1520)$, the $N^*$ resonances
make a negligible contribution to Compton scattering. The Compton
data are sensitive mainly to the $\Delta N\gamma$ couplings and the
$\sigma\gamma\gamma$ coupling. Since the $\Delta N\gamma$ parameters
are also strongly constrained by the pion photoproduction multipoles,
a simultaneous fit of both processes is not trivial: one essentially
has only the $\sigma\gamma\gamma$ coupling to adjust in fitting
Compton cross sections.

The model predicts the electic and magnetic polarizability of
the proton. At the real photon point, we obtain\footnote{The
polarizabilities are given in units 10$^{-4}$ fm$^3$.}:
$\alpha(0)+\beta(0)=8.23$
and $\alpha(0)-\beta(0)=3.05$. Hence, our result for the
electric polarizability is $\alpha(0)=5.64$, which is too small.
The magnetic polarizability is $\beta(0)=2.59$,
which agrees well with experiment~\cite{Mac95}: the
``global average'' data for the proton are
           $\alpha(0)=12.1(0.8)(0.5)$ and
           $\beta(0)=2.1(0.8)(0.5)$.
It is not obvious why the electric polarizability is underestimated.
This remains to be investigated, but it is likely related to the fact
that the modelling of two-pion exchange by a sharp $\sigma$-meson
is too crude. Also the fit to the Compton-scattering data leaves
room for improvement. In this connection we mention
calculations in Heavy Baryon ChPT where the polarizabilities
come from  pion-nucleon loops. The predicted values are:
$\alpha(0)=10.5(2.0)$, $\beta(0)=3.5(3.6)$~\cite{Ber93,Ber94}.

In summary, we have presented results for pion-nucleon scattering,
pion photoproduction, and Compton scattering using the $K$-matrix
formalism. The model used is fully relativistic, crossing symmetric,
and gauge invariant. Due to the simultaneous description of the three
different processes, our model correlates a huge amount of data,
resulting in a strongly constrained fit.
The four-star resonances below $\sqrt{s}\simeq 1.7$ GeV have been
included explicitly and the $\pi N^*N$ and $\gamma N^*N$ couplings
have been determined from a fit to the available partial-wave
amplitudes and cross sections.

\acknowledgements
\noindent
Part of this work was included in the research program
of the Stichting voor Fundamenteel Onderzoek der Materie
(FOM) with financial support from the Nederlandse Organisatie
voor Wetenschappelijk Onderzoek (NWO). A.Yu. K. acknowledges
a special grant from the NWO and would like to thank the
staff of the Kernfysisch Versneller Instituut for the kind
hospitality.

\begin{table}
\begin{tabular}{c|ccrr|rrr}
         & \multicolumn{2}{c}{$M_r$} & \multicolumn{2}{c|}{$\Gamma_0$} &
           \multicolumn{2}{c}{$g_{\pi N^*N}$} & \\
  $N^*$ resonance & Model & PDG & Model & PDG & Model & PDG & $a_\pi$ \\
\tableline
   $P_{11}(1440)$ & 1520 & 1440 & 372 & 230 & 6.272 & 5.142 & \\
   $D_{13}(1520)$ & 1505 & 1520 &  90 &  54 & 1.909 & 1.537 & $-0.910$ \\
   $S_{11}(1535)$ & 1535 & 1535 &  80 &  83 & 2.200 & 2.326 & \\
   $S_{11}(1650)$ & 1700 & 1650 &  95 &  45 & 3.630 & 2.375 & \\
   $P_{11}(1710)$ & 1710 & 1710 & 150 &  70 & 0.890 & 0.890 & \\
   $P_{33}(1232)$ & 1224 & 1232 & $-$ & $-$ & 2.050 & 2.116 & $-0.265$ \\
   $P_{33}(1600)$ & 1650 & 1600 & 200 & 290 & 0.301 & 0.501 & $-0.250$ \\
   $S_{31}(1620)$ & 1620 & 1620 & 112 & 112 & 3.641 & 2.319 & \\
   $D_{33}(1700)$ & 1650 & 1700 & 200 & 255 & 1.189 & 1.284 & $+0.280$ \\
\end{tabular}
\caption{Results for the different $N^*$ resonances: pole masses,
         one-pion couplings, two- (and multi-)pion widths,
         and off-shell parameters.}
\label{tab:1}
\end{table}

\begin{table}
\begin{tabular}{c|rrrrrr}
       & \multicolumn{2}{c}{$g_{1,\gamma N^*N}$} &
         \multicolumn{2}{c}{$g_{2,\gamma N^*N}$} &       &         \\
  $N^*$ resonance &  Model   & Ref.~\cite{Feu97} &
                     Model   & Ref.~\cite{Feu97} &
                                  $a_{1,\gamma}$ & $a_{2,\gamma}$  \\
\tableline
 $P_{11}(1440)$ & $-$0.97 &$-$0.400&        &        &        &      \\
 $D_{13}(1520)$ &    5.12 &  3.449 &   4.78 & 3.003  &$-$0.06 & 0.95 \\
 $S_{11}(1535)$ & $-$0.56 &$-$0.623&        &        &        &      \\
 $S_{11}(1650)$ &    0.59 &$-$0.205&        &        &        &      \\
 $P_{33}(1232)$ &    5.14 &  5.416 &   5.54 & 6.612  &$-$0.44 & 2.11 \\
 $S_{31}(1620)$ & $-$0.51 &  0.144 &        &        &        &      \\
 $D_{33}(1700)$ &    1.74 &  1.895 &   4.75 & 3.921  &   4.19 & 5.76 \\
\end{tabular}
\caption{Results for the different $N^*$ resonances:
         photon couplings, including off-shell parameters.
         For the $P_{11}(1710)$ and $P_{33}(1600)$ no photon
         couplings were fitted. The comparison is to ``Fit \#5''
         (resonance pole positions) of Ref.~\protect\cite{Feu97}.}
\label{tab:2}
\end{table}

\end{document}